\documentclass[preprint,journal]{vgtc}            % preprint (journal style)

\usepackage[frozencache,cachedir=.]{minted}

\setminted{
  fontsize=\normalsize,
  baselinestretch=1,
  breaklines=0  .
}

\usepackage{xargs}      % Used for new commands with optional arguments
\usepackage{soul}       % Used for custom comments
\usepackage{color}      % Used for custom colors in comments
\usepackage{xspace}     % Used for abbreviation spacing
\usepackage{xpunctuate} % Used for abbreviation spacing
\usepackage{multirow}
\usepackage{array}
\usepackage{multirow}
\usepackage{tabularx}
\usepackage{listings}
\usepackage{mdframed}
\usepackage{caption}
\newcommand{\etal}{{et~al\xperiod}\xspace}

\definecolor{lightpink}{RGB}{237,157,202}
\definecolor{lightred}{RGB}{210,121,121}
\definecolor{lightorange}{RGB}{230,170,50}
\definecolor{lightgold}{RGB}{210,194,121}
\definecolor{lightgreen}{RGB}{121,210,121}
\definecolor{lightaqua}{RGB}{121,206,210}
\definecolor{lightblue}{RGB}{121,124,210}
\definecolor{lightpurple}{RGB}{153,102,255}
\definecolor{red}{RGB}{178,34,34}
\definecolor{gray}{RGB}{166,166,166}
\definecolor{forestgreen}{RGB}{74,103,65}

\newcommand{\add}[1]{#1}
\newcommand{\cut}[1]{\textcolor{red}{}}

\newcommandx{\guest}[3][1=]
    {\setulcolor{lightorange}{\ul{#1}} \textcolor{lightorange} %% Usage: \guest[Underline]{Name}{Comment}
    {[\textbf{#2:} #3]}}
\newcommandx{\leilani}[2][1=] 
    {\setulcolor{violet}{\ul{#1}} \textcolor{violet}
    {[\textbf{Leilani:} #2]}}
\newcommandx{\jeff}[2][1=] 
    {\setulcolor{lightgold}{\ul{#1}} \textcolor{lightgold}
    {[\textbf{Jeff:} #2]}}
\newcommandx{\will}[2][1=] 
    {\setulcolor{lightgreen}{\ul{#1}} \textcolor{lightgreen} 
    {[\textbf{Will:} #2]}}    
\newcommandx{\mlg}[2][1=] 
    {\setulcolor{lighblue}{\ul{#1}} \textcolor{lightblue}
    {[\textbf{Mitchell:} #2]}}

\onlineid{1472}

\vgtccategory{Research}

\title{DracoGPT: Extracting Visualization Design Preferences\\ from Large Language Models}

\author{%
  Huichen Will Wang,
  Mitchell Gordon,  
  Leilani Battle, and
  Jeffrey Heer
}

\authorfooter{
  \item
  	All authors are with the University of Washington.
  	E-mail: \{wwill, mgord, leibatt, jheer\}@cs.washington.edu.
}

\abstract{%
  Trained on vast corpora, Large Language Models (LLMs) have the potential to encode visualization design knowledge and best practices. However, if they fail to do so, they might provide unreliable visualization recommendations. What visualization design preferences, then, have LLMs learned? We contribute DracoGPT, a method for extracting, modeling, and assessing visualization design preferences from LLMs. To assess varied tasks, we develop two pipelines---DracoGPT-Rank and DracoGPT-Recommend---to model LLMs prompted to either rank or recommend visual encoding specifications. We use Draco as a shared knowledge base in which to represent LLM design preferences and compare them to best practices from empirical research. We demonstrate that DracoGPT can accurately model the preferences expressed by LLMs, enabling analysis in terms of Draco design constraints. Across a suite of backing LLMs, we find that DracoGPT-Rank and DracoGPT-Recommend moderately agree with each other, but both substantially diverge from guidelines drawn from human subjects experiments. Future work can build on our approach to expand Draco's knowledge base to model a richer set of preferences and to provide a robust and cost-effective stand-in for LLMs.
}


\graphicspath{{figs/}{figures/}{pictures/}{images/}{./}} % where to search for the images

\usepackage{tabu}                      % only used for the table example
\usepackage{booktabs}                  % only used for the table example
\usepackage{lipsum}                    % used to generate placeholder text
\usepackage{mwe}                       % used to generate placeholder figures

\usepackage{mathptmx}                  % use matching math font

\begin{document}

%%%%%%%%%%%%%%%%%%%%%%%%%%%%%%%%%%%%%%%%%%%%%%%%%%%%%%%%%%%%%%%%
%%%%%%%%%%%%%%%%%%%%%% START OF THE PAPER %%%%%%%%%%%%%%%%%%%%%%
%%%%%%%%%%%%%%%%%%%%%%%%%%%%%%%%%%%%%%%%%%%%%%%%%%%%%%%%%%%%%%%%


\firstsection{Introduction}
\maketitle

Large language models (LLMs) have shown potential for visualization tasks including captioning~\cite{Tang2023VisTextAB, huang2023lvlms},  generation~\cite{maddigan2023chat2vis}, and critique~\cite{kim2023good}. State-of-the-art LLMs are trained on vast corpora of texts or even images, which likely include a wide range of visualization discussions, research papers, and code examples. Hence, LLMs might also encode visualization design knowledge and best practices. What visualization design preferences, then, have LLMs learned? As LLM recommendations may influence people's analysis and decision-making, poor decisions could stem from suboptimal LLM visualization designs. \add{To the best of our knowledge, no work has yet proposed methods to systematically understand LLMs' visualization design preferences, partly due to the difficulties in eliciting and representing these preferences.}

\add{We contribute \emph{DracoGPT}, a method for extracting, modeling, and assessing visualization design preferences from LLMs.} Using LLM responses to generate training data\cut{ for a Draco knowledge base}, DracoGPT synthesizes comparable Draco knowledge bases~\cite{moritz2018formalizing, yang2023draco} to model and assess LLM preferences.
The DracoGPT method may be applied across various LLMs, prompts, and tasks in order to assess, compare, and reuse design preferences.
By comparing Draco knowledge bases fit to LLM responses to those fit to human performance data, we can interrogate the extent to which LLM preferences align with best practices from empirical visualization research, and clarify how they diverge.

\add{DracoGPT relies on Draco to represent visualizations as a set of \textit{facts} and} constructs a knowledge base in terms of \emph{logical constraints} over encoding specifications\add{, where each constraint corresponds to a design choice.} Design preferences are expressed as numerical weights for soft constraints, which are learned from training data in the form of \emph{visualization pairs} in which one chart is \add{preferred over}\cut{deemed preferable to} the other. Prior work~\cite{moritz2018formalizing, zeng2023too} learns Draco constraint weights using chart pairs from experimental research, such that subsequent chart scoring and recommendation adheres to empirically-derived visualization best practices.

To extract design preferences, we prompt LLMs \add{using two tasks:} directly labeling chart pairs (\emph{rank}) or synthesizing (\emph{recommend}) new charts that we can then contrast with alternatives.
\cut{Both processes produce visualization pairs from which we can learn Draco constraint weights.}
As an initial assessment of \emph{task} differences on expressed LLM preferences, we contribute two pipelines: DracoGPT-Rank uses the \emph{discriminative} task of selecting a preferred chart, while DracoGPT-Recommend uses the \emph{generative} task of \add{synthesizing chart completions from partial specifications}.

We further conduct a case study with DracoGPT comparing LLM ranking and recommendation preferences with human performance data.
We use experimental stimuli from Kim et al.~\cite{kim2018assessing}\add{, which include variations of three-variable scatterplot designs for both value comparison and summary chart reading tasks,} and test three LLMs (GPT3.5-Turbo, GPT4, and GPT4-Turbo). We find that DracoGPT-Rank and -Recommend are able to learn knowledge base configurations whose preferences accurately match LLM judgments, \cut{ (e.g., test set accuracy of 99\% (rank) and 97\% (recommend) using GPT4-Turbo). We can} allowing us to analyze LLM preferences by comparing fitted Draco knowledge bases.
\cut{For the ranking task, we find that GPT4-Turbo exhibits some preferences akin to those observed by Kim et al.~(such as the tendency to encode the most important quantitative variable in the chart with \texttt{x} rather than \texttt{size} or \texttt{color}), yet many of the model's preferences diverge from experimental results.}
\cut{For example, with charts intended to convey summary statistics, GPT4-Turbo strongly penalizes a continuous \texttt{size} encoding, in contrast to Kim~\etal's experimental results.}
Draco chart costs derived from GPT4-Turbo \textit{rank} and \textit{recommend} pipelines moderately correlate\cut{($r$ = 0.70)} with each other, while both correlate weakly with costs learned from Kim et al.'s results\cut{($r$ = 0.36 for rank, $r$ = 0.40 for recommend)}, thus diverging from empirical performance data.
\add{More specifically, we observe that GPT4-Turbo's preferences moderately align with Kim~\etal's findings for perceptual tasks involving the comparison of individual values\cut{($r$ = 0.69 for rank, $r$ = 0.67 for recommend)}, but do not align for aggregate summary tasks.}
Meanwhile, GPT4-Turbo expresses nearly identical preferences\cut{($r$ = 0.99)} when recommending charts as Vega-Lite JSON~\cite{satyanarayan2016vega} or Vega-Altair code~\cite{vanderplas2018Altair}, exhibiting consistency across these different (though highly similar) tools.

DracoGPT provides a method to systematically test how prompting and chart representations affect LLM design preferences. It can readily be applied to study other LLMs or preference elicitation methods. \cut{In the future, Draco models could also serve as a robust and cost-effective stand-in for LLMs, whereas LLMs can help identify expressivity limits of Draco's constraints and might expand knowledge bases to encode a richer space of preferences.} In sum,
our primary contributions are:
\begin{enumerate}
    \item DracoGPT, a method for extracting, modeling, \add{and assessing} visualization preferences from LLMs by eliciting training data and learning design preferences in a Draco knowledge base. \add{We employ four strategies---examining training pairs, soft constraint counts, soft constraint weights, and correlation of summed chart costs---to analyze the preferences of fitted Draco models};
    \item A case study using three LLMs and stimuli from Kim et al.~\cite{kim2018assessing}, demonstrating DracoGPT's ability to accurately model LLM judgments and enable comparison of preference models from both LLMs and results of human subjects experiments.
    
\end{enumerate}

\begin{figure*}[t!]
    \centering
    \includegraphics[width=\textwidth]{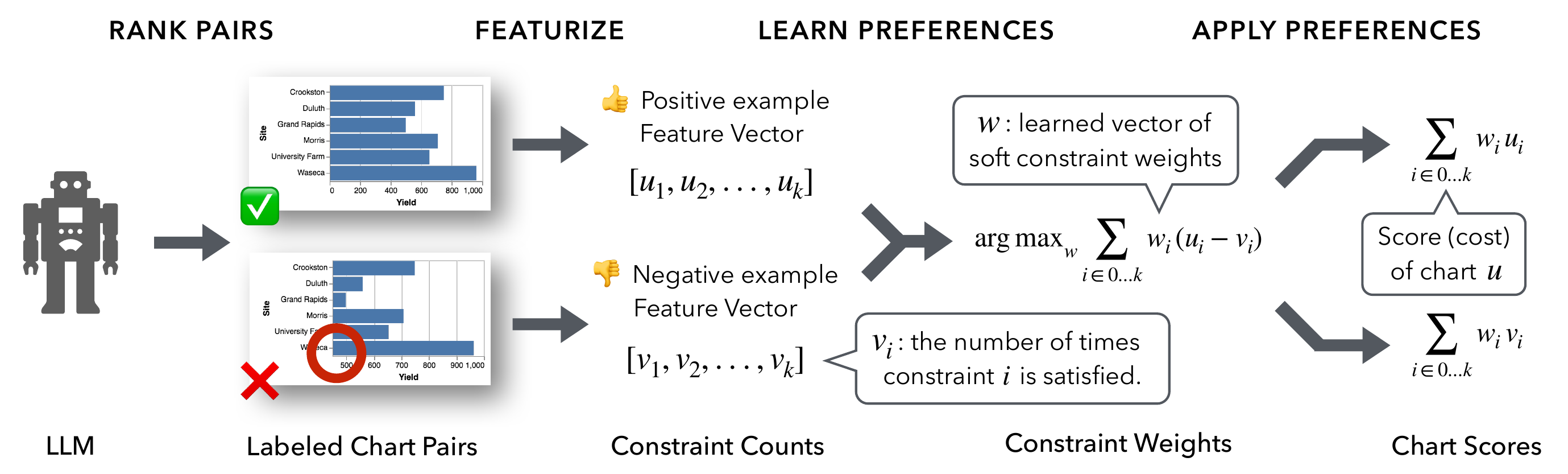}
    \vspace{-0.5cm}
    \caption{Overview of the DracoGPT-Rank pipeline. (1) \add{User provides prompt templates for} an LLM to rank chart pairs; (2) Draco featurizes charts and produces feature vectors consisting of constraint counts; (3) Draco learns constraint weights over \add{LLM-labeled chart}\cut{a corpus of labeled} pairs by fitting a RankSVM model; (4) The fitted Draco model can be applied to score charts. Results at each stage of the pipeline afford insight into LLM ranking preferences.}
    \label{fig:rank diagram}
    \vspace{-0.3cm}
\end{figure*}

\section{Related Work}

\subsection{Graphical Perception Knowledge in VizRec Systems}
Creating a visualization entails various design decisions, each having potentially significant implications how reader perceive and interpret the visualization~\cite{borkin2015beyond, cleveland1984graphical, kim2018assessing, stokes2023role}.\cut{on factors like reader memorability, speed and accuracy of chart interpretation, key takeaways, and biases} To simplify this process, researchers have developed various visualization recommendation (VizRec) systems~\cite{zhu2020survey}. Though many such systems achieve the goal of automating visualization creation, there are no guarantees on how effective the recommended charts are. In fact, \cut{according to a survey on VizRec systems by Zeng et al.,} existing systems severely under-utilize the literature of graphical perception research, with no one system synthesizing results from more than three graphical perception studies~\cite{zeng2023review}.

Synthesizing graphical perception literature into a VizRec system is non-trivial. Traditional visualization recommendation approaches use rule-based systems, machine learning~\cite{wang2023llm4vis}, or a mix of both. Rule-based systems~\cite{mackinlay1986automating, wongsuphasawat2017voyager} rely on the VizRec system designer authoring rules that reflect best practices established in the literature. Thus, integrating new graphical perception research often necessitates adding more rules, a process that can be labor-intensive. Additionally, as the rule set expands, ensuring seamless rule integration becomes increasingly challenging \add{given the possible interplay and contradictions between rules}. Machine-learning-based approaches (e.g.,~\cite{hu2019vizml, li2021kg4vis, zhou2021table2charts}), on the other hand, are trained on curated datasets. These systems typically need training data in various formats, ranging from chart pairs featuring a positive example and a negative example~\cite{moritz2018formalizing} to dataset-visualization pairs~\cite{hu2019vizml, li2021kg4vis}. However, existing graphical perception papers rarely publish all stimuli used or share stimuli in a format that can be readily utilized to train machine-learning-based systems, limiting their uptake by VizRec systems. \add{Moreover, it is often challenging to assess how well the trained models encode graphical perception principles, hence the necessity to develop probes to extract visualization design preferences. In our case study, we use stimuli from Kim~\etal~\cite{kim2018assessing} to probe LLMs' design preferences, which cover a diverse range of visualization task types that involve reasoning over both individual and aggregate summary values.}

\subsection{LLMs for Visualization}
LLMs demonstrate impressive capabilities across many natural language processing tasks~\cite{bubeck2023sparks} and have made inroads into data analysis~\cite{cheng2023is, ma2023insightpilot, mcnutt2023design, weng2024insightlens, Gu2023HowDA, Li2023WhereAW}. The Internet-scale training data behind LLMs presumably includes rich information on visualization. As such, many researchers are exploring the potential of LLMs for visualization tasks. For example, Tang et al.~\cite{Tang2023VisTextAB} and Huang et al.~\cite{huang2023lvlms} explore caption generation for visualizations with LLMs. \add{Ko~\etal~\cite{Ko2023NaturalLD} further demonstrates the potential of LLMs in generating rich and diverse natural language datasets for visualizations. Another vein of research harnesses the power of LLMs in interpreting, storing, and summarizing information to facilitate visualization reading~\cite{zhao2024leva, choe2024enhancing}. In addition to generating text based on visualizations, LLMs also provide a convenient interface for natural language to visualization (NL2VIS), outperforming baseline methods across a suite of benchmarks~\cite{li2024visualization, wu2024automated, tian2024chartgpt, luo2021synthesizing, Vaithilingam2024DynaVisDS}.} 

\add{While previous work has explored using LLMs to generate visualizations, no studies have yet proposed methods to understand the visualization design preferences of LLMs.} Most similar to our work is a study comparing LLMs to humans for answering visualization-related questions~\cite{kim2023good}. The authors compiled a dataset of 119 visualization design questions for both humans and LLMs to answer. Through qualitative coding, they discovered that ChatGPT's responses were better in terms of breadth, clarity, and coverage. 
Nonetheless, this approach does not characterize specific design preferences expressed by ChatGPT, nor does it compare how ChatGPT's preferences diverge from best practices. Our work provides an automated and scalable solution to this problem, quantitatively models LLM visualization design preferences for different tasks, and allows for direct comparison of design preferences across sources (i.e., different LLMs and empirical results).

\section{Extracting Design Preferences with DracoGPT}
\add{In this section, we provide an overview of Draco and how the DracoGPT method extracts, models, and assesses visualization design preferences of LLMs across two tasks: ranking and recommending visualizations.}

\add{\subsection{Overview of Draco}}
We model visualization design preferences in the context of Draco~\cite{moritz2018formalizing, yang2023draco}, a reasoning and recommendation system based on logic programming methods.
\add{The fundamental building blocks of visualization representation in Draco are chart \emph{facts}, which capture properties of the visualized data and the visualization specification. For instance, ``a mark in the visualization uses the \texttt{x} channel'' is a fact Draco can encode. To model the complex visualization design space, Draco composes lower-level \emph{facts} into} logical statements, subject to logical \emph{constraints}. \add{Collectively, these constraints constitute Draco's \emph{knowledge base}, which is the set of rules and guidelines it uses to validate and evaluate visualizations.}
Draco's knowledge base contains both hard and soft constraints. Hard constraints, such as ``line or area marks require both \texttt{x} and \texttt{y} channels'', must be respected to avoid ill-formed or non-expressive charts. Soft constraints, on the other hand, include a numerical \emph{weight} to represent design preferences in terms of cost trade-offs: soft constraints incur a penalty (positive weight) or reward (negative weight) when satisfied.

Draco then treats visualization recommendation as a constrained optimization problem, searching and ranking a space of feasible charts.
The \emph{cost} (or score) of a chart in Draco is the sum of the weights of all satisfied soft constraints.
For example, \texttt{linear\_color} (use of a linear \texttt{color} scale) and \texttt{ordinal\_size} (use of an ordinal \texttt{size} scale) are two soft constraints that Draco might consider.  
Given a query in the form of a partial chart specification\add{, i.e., some facts about the visualization}, Draco enumerates a design space to find complete specifications that minimize these summed weights\add{, which it then renders using visualization tools like Vega-Lite}.

To learn constraint weights from data and thereby create a new Draco instance, one provides a dataset of annotated chart pairs, with a positive example deemed a ``better'' design than a negative example. For each chart, Draco counts how many times each soft constraint is satisfied, representing each chart as a feature vector of these soft constraint counts (i.e., one dimension per soft constraint). Draco then trains a linear RankSVM model to distinguish the vectors for the positive examples from those of the negative examples. The learned RankSVM model parameters are then the fitted Draco soft constraint weights that best discriminate positive from negative examples.

\begin{figure*}[ht!]
    \centering
    \includegraphics[width=\textwidth]{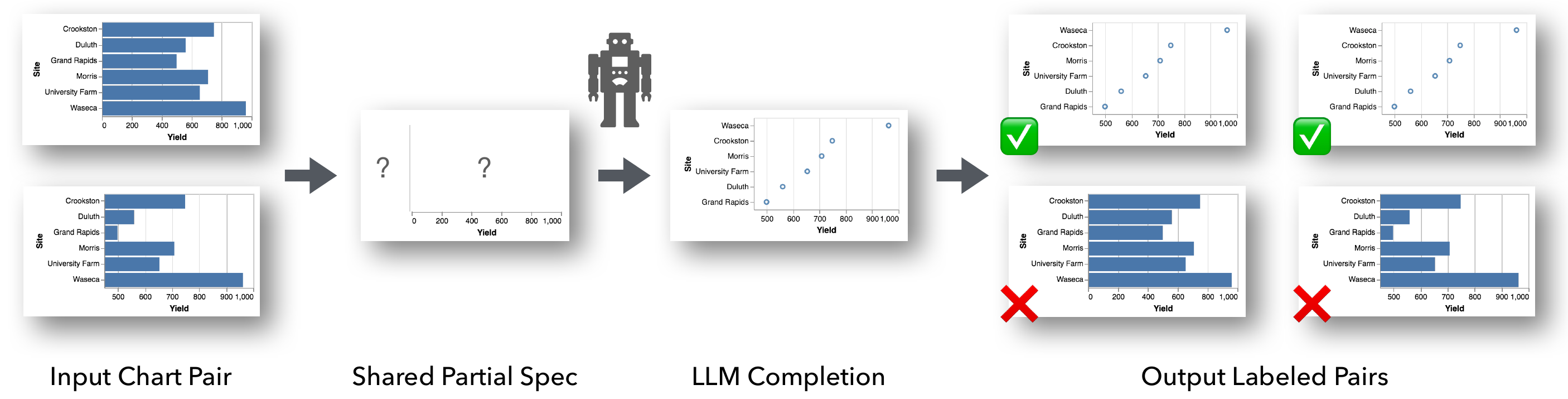}
    \vspace{-0.5cm}
    \caption{Overview of the DracoGPT-Recommend chart pair construction pipeline. (1) Given an input chart pair, \add{the pipeline extracts} their shared partial specification, then (2) prompts an LLM to ``optimally'' complete the partial specification. (3) \add{The pipeline constructs} up to two new chart pairs for training a Draco model: the LLM completion is labeled as the positive example and an input chart as the negative example.}
    \label{fig:rec diagram}
    \vspace{-0.3cm}
\end{figure*}

\subsection{Two Visualization Tasks: Rank and Recommend}
Consider two tasks: ranking visualizations according to their effectiveness (e.g., as measured by affordance for reading speed and accuracy\cut{, or other behavioral measures}) and recommending a full visualization specification given a partial specification.
The ranking task is \emph{discriminative}, requiring a preference judgment over a fixed set of options.
The recommendation task is \emph{generative}, and requires synthesizing new candidate charts.
Both tasks involve design preferences, are relevant to visualization creators, and central to many VizRec systems~\cite{zeng2023review,zeng2023too}.

Draco, for instance, can perform both of these tasks. Given an input visualization, it can compute a total cost for it by tallying the weights of all satisfied constraints. As such, ranking two visualizations reduces to comparing their costs. When recommending a visualization given a partial specification, Draco enumerates possible completions and draws on the same knowledge base to score them to determine top recommendations. Thus, Draco is an example of a \textit{consistent} system, one that applies the same set of visualization design preferences when ranking and recommending. If a VizRec system encodes design best practices, then consistency is a desirable property, since it guarantees that shared best practices are observed across tasks.

Draco operates on a linear RankSVM model with about 150 parameters, while state-of-the-art LLMs possess billions of parameters and perform stochastic decoding~\cite{brown2020language, touvron2023llama2}.
While we can prompt LLMs to serve as visualization rankers and recommenders, they may not be consistent.
Across the spectrum of visualization tasks LLMs can perform, differents tasks might elicit different conditional probability distributions, resulting in varied design preferences.

In response, we develop two complementary pipelines to initially test such sensitivities: DracoGPT-Rank, for modeling design preferences when LLMs discriminatively rank charts, and DracoGPT-Recommend, for modeling preferences when they generatively recommend charts. Using Draco as a common knowledge base, we can systematically probe and compare visualization design preferences demonstrated by LLMs across tasks, as well as best practices from empirical research. We now present each pipeline in detail.

\subsection{DracoGPT-Rank: Extracting LLM Design Preferences when Ranking Charts}
\label{rank procedure}

In order to train DracoGPT-Rank, we need to prepare a dataset of chart pairs. In practice, such a dataset is typically collected from the literature and the labels are derived using theoretical analyses or empirical results~\cite{zeng2023too}. In order to extract visualization design preferences from LLMs when they are ranking charts, we can provide chart pairs without labels and instruct an LLM to select the ``better'' design. As such, an existing corpus of chart pairs can be used to generate training data and fit new Draco models, as illustrated in Figure~\ref{fig:rank diagram}. The chart pairs to be fed to LLMs can be specified in any format (e.g., Vega-Lite~\cite{satyanarayan2016vega}, ggplot2~\cite{wickham2016ggplot2}, or---in the case of multimodal models---potentially even bitmap images), so long as they can be recognized by the LLMs and transformed into a logical representation compatible with Draco.

A potential complication is that LLMs are sensitive to positional bias. For example, LLMs often favor the first and last choices when answering multiple choice questions~\cite{pezeshkpour2023large}. This sensitivity to the ordering of items is similarly observed when employing LLMs as recommendation systems~\cite{ma2023large}. To mitigate positional bias, \add{the pipeline} shuffles the order within each chart pair and separately prompts models to rank either arrangement. If the rankings conflict between two orderings of the same chart pair, \add{the pipeline} excludes the pair from the training data, as the inability to consistently determine rankings likely suggests that noise (positional bias) dominates signal (visualization design preferences). 

\add{This process leads to a dataset to train DracoGPT-Rank. At this point, we can analyze the design preferences of LLMs by comparing the design choices in positive and negative charts. Our pipeline further} splits this dataset into training and test sets and performs cross-validation to identify optimal hyperparameters. Then, it trains a DracoGPT-Rank model on the entire training set and assesses accuracy on the test set,\cut{Since the test set is not used to train the model, The accuracy on the test set} which allows an unbiased evaluation of how well DracoGPT-Rank fits to LLM-provided labels. If this accuracy is reasonably high, we have some assurance that the fitted DracoGPT-Rank instance models the LLM's preferences well. We can then use the optimal hyperparameters identified in cross-validation to train on the entire dataset and analyze the resulting DracoGPT-Rank model\add{, such as by comparing the number of times each constraint is satisfied in positive and negative charts and by inspecting the weights for each constraint~(\S\ref{rank preference})}.

\subsection{DracoGPT-Recommend: Extracting LLM Design Preferences when Recommending Charts}
\label{rec procedure}

Visualization recommendation is arguably more widely undertaken with LLMs than ranking. Users can specify to LLMs what data they want to visualize and (optionally) note additional design details they want the recommendations to respect.
To produce training data that captures LLM recommendation preferences, one strategy is to generate partial specifications of visualizations from scratch and ask for optimal completions from LLMs, which then serve as positive examples in chart pairs.
Partial specifications can be in any format, so long as they can be productively processed by the LLM.
Similarly, completions can be in any reasonable visualization specification format, provided they can be mapped to Draco's logical format. We show how Vega-Lite and Vega-Altair can be transformed into Draco's chart representation Domain Specific Language (DSL) in our case study in \S\ref{recommend case study}. To construct negative examples, one method is to sample valid completions of the partial specification that differ from the LLM's recommendation. Since we request optimal completions from the LLM, any alternative completion is implicitly deemed inferior by the LLM. 

Alternatively, we can use existing chart pairs to generate training data for DracoGPT-Recommend, as shown in Figure~\ref{fig:rec diagram}. Given a chart pair, we can programmatically extract the common parts of their specifications. 
Then, we can prompt an LLM to provide completions, which act as positive examples. Since both charts in the original chart pair are valid completions of the common parts of their specifications, they can both serve as negative examples if they differ from the LLM's provided completion.
Once the training set is constructed, we can train and analyze DracoGPT-Recommend in the same manner as DracoGPT-Rank.

\section{Case Study using DracoGPT-Rank}

To \add{showcase} the DracoGPT approach and perform an initial analysis of LLM-expressed design preferences, we present a case study using a corpus of charts developed as experimental stimuli by Kim~\etal~\cite{kim2018assessing}.
\add{We begin by training and analyzing DracoGPT-Rank models to understand LLM ranking preferences and compare them to empirical findings.}

\label{rank case study}

\subsection{Dataset}

\add{Despite covering only one chart type (scatterplots),} our choice of dataset includes a variety of encodings and backing data variables, is compatible with both Draco and Vega-Lite, and provides human performance data that can serve as a baseline for comparison.
Kim~\etal~\cite{kim2018assessing} experimentally assessed the effects of visual encoding channels, task type, and data distributions on the effectiveness of data visualizations.
Their visual stimuli each depict three data variables---one categorical (\texttt{n}) and two quantitative (\texttt{q1} and \texttt{q2})---and employ a total of 12 encoding specifications that include \texttt{x}, \texttt{y}, \texttt{color}, \texttt{size}, and \texttt{row} (faceting) channels.
Using these 12 designs, they varied the data distribution (i.e., the number of data points, cardinality of the categorical variable, and entropies of the quantitative variables).
They then conducted crowdsourced experiments measuring participants' performance in terms of reading speed and accuracy across two task categories: \emph{value tasks} (reading or comparing individual visual marks) and \emph{summary tasks} (identifying or comparing aggregate properties of visual marks).
For value tasks, participants answered questions about \texttt{q1}; in summary tasks, they answered questions about \texttt{q1} and \texttt{n}. These variables are referred to as ``variables of interest''.

To train a Draco knowledge base, Moritz et al.~\cite{moritz2018formalizing} constructed 1,152 chart pairs from Kim et al.'s stimuli by identifying all visualization pairs that depict the same data and correspond to statistically significant differences in reading speed or accuracy.
The charts associated with better performance are labeled as positive examples.
As outlined in \S\ref{rank procedure}, for DracoGPT-Rank we permute the order within each of the 1,152 chart pairs, resulting in 2,304 chart pairs to be passed to LLMs. Using this dataset as a probe, we investigate how LLM ranking preferences compare with Kim et al.'s experimental results. 

\subsection{Chart Specification}
\label{sec:chart-spec}

Charts are represented in Draco as a set of logical facts---including facts about the data depicted and the visualization design---using Answer Set Programming (ASP) syntax.
One drawback of the ASP representation is its limited readability.
In our LLM prompts we transform these facts into Draco's dictionary format, which is as expressive as the ASP specification and closely resembles Vega-Lite. For instance, the following snippet, here in YAML format, indicates a (potentially partial) visualization specification in which the data field \texttt{q1} is encoded in the \texttt{x} channel using a \texttt{linear} scale that includes a \texttt{zero} baseline.

\vspace{-0.25cm}
\begin{minted}{yaml}
view:
- mark:
  - type: point
    encoding:
    - { channel: x, field: q1 }
- scale:
  - { channel: x, type: linear, zero: true }
\end{minted}

\subsection{Experiment Setup}

We conduct experiments with three LLMs from OpenAI: GPT-4-0125-preview (hereafter GPT4-Turbo), GPT-4-0613 (hereafter GPT4), and GPT-3.5-Turbo-0125 (hereafter GPT-3.5-Turbo) to demonstrate that our approach is applicable across models. \add{We select these models because they are popular, state-of-the-art LLMs at the time of writing.} In order to elicit stable visualization design preferences, we set the decoding temperature to zero to ensure minimal stochasticity. To train DracoGPT-Rank, we partition the dataset, allocating 80\% for training and the remaining 20\% for testing. Furthermore, we employ 5-fold cross-validation to conduct a grid search, aimed at identifying an optimal regularization parameter. 

\subsection{Prompts}

Our LLM prompts introduce the ranking task, describe fields in a chart specification, and outline the expected format for the response.
The prompts specifically indicate to rank charts for either a value comparison or summary task, following the logic of Kim et al. 
We designed our prompts to be relatively simple, incorporating only the essential information required for the tasks and abstaining from other strategies, such as Chain of Thought prompting~\cite{wei2022chain}.
To provide appropriate context, our prompts include guidance to ``rely upon visualization design best practices and graphical perception research to rank visualizations.''
In early pilot experiments, including this instruction appeared to meaningfully affect the results.
We performed initial experiments using several chart pairs to ensure a high degree of instruction following.
The supplemental materials include an example ranking prompt.

Although we acknowledge that subtle changes to textual prompts can sometimes lead to different LLM responses, we believe our prompts to be a reasonable starting point for demonstrating the DracoGPT pipeline and eliciting implicit design preferences from LLMs.
Leveraging our pipeline, future work can further investigate the effects of prompting strategies on the expressed design preferences.

\subsection{Fitted Draco Models}
\label{rank results}

For each chart pair, we check the consistency of LLM responses across the two presentation orders. As we pass chart pairs in two orders, inconsistent responses for a chart pair are either both ``Chart 1'' or both ``Chart 2''. As shown in Table~\ref{tab:accuracy}, of the three LLMs tested, GPT4-Turbo shows the lowest percentage of inconsistent responses at 23.09\%, followed by GPT4 at 28.56\%. For GPT3.5-Turbo, the vast majority (72.48\%) of its responses are conflicting. We observe that \textbf{even state-of-the-art LLMs often fail to provide consistent answers to the same ranking question.} We further examine the distribution of choices for the inconsistent responses. As shown in Table~\ref{tab:chart_conflicting}, over 90\% of the inconsistent responses are ``Chart 1'' for GPT4 and GPT3.5-Turbo, signifying a strong positional bias for the first choice. This tendency is weaker in GPT4-Turbo, which selects ``Chart 2'' 62.40\% of the time. \add{One possible explanation for inconsistent LLM responses across chart pair orders is that when both charts in a pair are evaluated roughly equally by the LLM, the response is driven by other, non-design-related biases such as a positional bias.}

We use only the consistently labeled chart pairs to train DracoGPT-Rank (i.e., the data in the first two columns of Table~\ref{tab:accuracy}).  
% Table~\ref{tab:accuracy} details the proportions of chart pairs whose LLM-provided labels agree or disagree with results from Kim et al.
Of all chart pairs, 63.54\% of GPT4-Turbo responses, 56.51\% of GPT4 responses, and a meager 9.98\% of GPT3.5-Turbo responses match the labels of the Kim et al. data. This initial result already indicates that \textbf{LLM chart rankings diverge from experimental results}.

\begin{table}[t!]    
    
  \scriptsize
  \centering
  \begin{tabu} to \linewidth {@{} X[1,l] *{3}{X[c]} @{}}
    \toprule
    & {Agree with} & {Disagree with} & {Inconsistent} \\
    & {Kim et al.} & {Kim et al.} & {Responses} \\
    \midrule
    GPT4-Turbo & 63.54\% & 13.37\% & 23.09\% \\
    GPT4 & 56.51\% & 14.93\% & 28.56\% \\
    GPT3.5-Turbo & 9.98\% & 17.53\% & 72.48\% \\
    \bottomrule
  \end{tabu}
  \vspace{-0.2cm}
  \caption{Percentages of LLM rankings that agree or disagree with Kim et al.'s data~\cite{kim2018assessing}, or are self-conflicting across the two pair orders. Pairs with inconsistent LLM responses are excluded from the training data.}
  \label{tab:accuracy}
  % Chart pairs for which an LLM provides inconsistent responses between the two orders are excluded from the DracoGPT-Rank training data for that LLM.
  
%   \end{table}

% \begin{table}[t]
  \vspace{0.3cm}
  \scriptsize
  \centering
  \begin{tabu}{l*{2}{c}}
    \toprule
    & {Both ``Chart 1''} & {Both  ``Chart 2''} \\
    \midrule
    GPT4-Turbo & 37.59\% & 62.40\% \\
    GPT4 & 92.71\% & 7.29\% \\
    GPT3.5-Turbo & 96.53\% & 3.47\% \\
    \bottomrule
  \end{tabu}
  \vspace{-0.2cm}
  \caption{Breakdown of inconsistent responses, as the percentages of LLM choices that are both ``Chart 1'' or both ``Chart 2''. GPT4 and GPT3.5-Turbo exhibit strong positional biases in favor of the first option.}
  \label{tab:chart_conflicting}
% \end{table}

% \begin{table}[t]
  \vspace{0.3cm}
  \scriptsize
  \centering
  \begin{tabu}{l*{2}{c}}
    \toprule
    & {Average CV} & {Test Set} \\
    & {Accuracy} & {Accuracy} \\
    \midrule
    GPT4-Turbo Rank & 98.87\% & 99.44\% \\
    Kim et al. Data (GPT4-Turbo Subset) & 94.77\% & 96.07\% \\
    \midrule
    GPT4 Rank & 99.39\% & 99.39\% \\
    Kim et al. Data (GPT4 Subset) & 95.44\% & 94.55\% \\
    \midrule
    GPT3.5-Turbo Rank & 97.64\% & 96.88\% \\
    Kim et al. Data (GPT3.5-Turbo Subset) & 89.33\% & 95.31\% \\
    \midrule
    Kim et al. Data (Full) & 93.81\% & 93.94\% \\
    \bottomrule
  \end{tabu}
  \vspace{-0.2cm}
  \caption{Average cross-validation and test accuracies for the three instances of DracoGPT-Rank and four instances of Draco trained on experimental human performance data from Kim et al.}
  \label{tab:rank acc}
  \vspace{-0.5cm}
\end{table}

We train DracoGPT-Rank models for each LLM. To compare LLM preferences against empirically established best practices, for each DracoGPT-Rank model we train an instance of Draco on the same data subset (consistent pairs only), but using the labels from Kim et al.'s experimental results.
In addition, we train an instance of Draco on the full dataset from Kim et al.
Table~\ref{tab:rank acc} shows the average 5-fold cross-validation (CV) accuracy and test set accuracy for these seven models.
All Draco instances fit their chart pair labels well, with test set accuracy well above 90\%.
These high accuracies confirm that our \textbf{DracoGPT-Rank instances model the design preferences of their source LLMs well}. Therefore, we proceed to investigate LLM design preferences by probing these DracoGPT-Rank models.

Interestingly, all DracoGPT-Rank models achieve higher test set accuracy than Draco models trained on the same pairs with labels from Kim et al. As Draco uses linear RankSVM to distinguish between positive and negative examples, this result indicates that, relative to Draco's knowledge base, the LLM ranking outputs are more linearly separable than Kim et al.'s experimental data. In other words, LLM design preferences may be more accurately modeled as Draco soft constraint weights than those synthesized from empirical studies.

\begin{figure}[t!]
    \centering
    \includegraphics[width=\columnwidth]{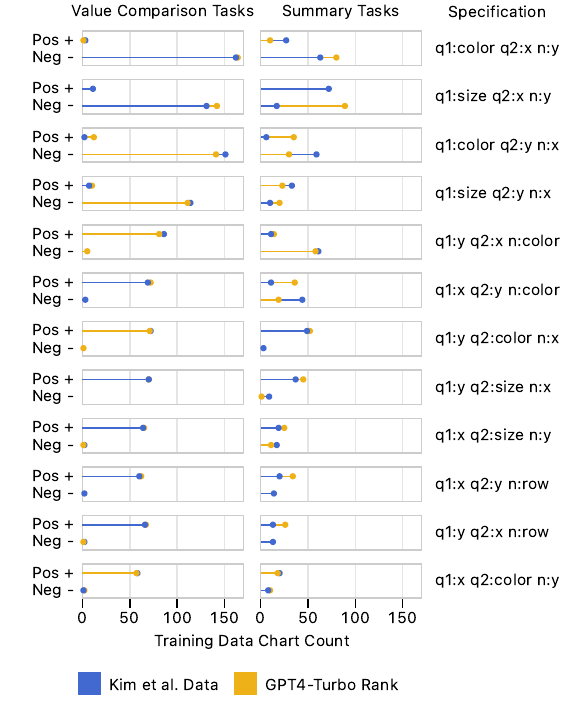}
    \vspace{-0.6cm}
    \caption{Distribution of positive and negative examples by encoding specification and interpretation task type for Kim~\etal and GPT4-Turbo. Only chart pairs for which GPT4-Turbo provides consistent responses are included. These training sets have similar distributions for value tasks, but notably diverge across summary tasks.}
    \label{fig:agg pos and neg examples}
    \vspace{-0.4cm}
\end{figure}

\subsection{Design Preferences in DracoGPT-Rank}
\label{rank preference}

We next analyze LLM preferences modeled by DracoGPT-Rank and compare them against results from Kim et al.
For clarity, we focus on Draco instances fit to GPT4-Turbo responses, as GPT4-Turbo is OpenAI's most recent state-of-the-art LLM at the time of writing and exhibits the highest test set accuracy.
We first examine chart pairs for which an LLM provides consistent responses and investigate what encoding specifications are preferred by LLMs. We then take a closer look at how LLM preferences and Kim et al. results diverge in terms of soft constraint counts. Finally, we examine the weights associated with soft constraints in the fitted DracoGPT model.

\subsubsection{LLMs' Preferred Encoding Choices}
\label{rank enc}
Selecting the appropriate encoding channel for each field is a critical design choice in the creation of visualizations.
We examine chart pairs for which GPT4-Turbo provides consistent responses and count how often each encoding specification is labeled as positive or negative according to GPT4-Turbo and Kim et al. 
Figure~\ref{fig:agg pos and neg examples} shows the distribution of these examples by encodings and task type (summary or value). 

Figure~\ref{fig:agg pos and neg examples} reveals that GPT4-Turbo generally prefers positional encodings for the most important quantitative variable, \texttt{q1}. For instance, it strongly favors \texttt{q1:y~q2:x~n:row} (encoding \texttt{q1} with \texttt{y}, \texttt{q2} with \texttt{x}, and encoding \texttt{n} with \texttt{row}), \texttt{q1:y~q2:size~n:x}, \texttt{q1:y~q2:color~n:x}, and \texttt{q1:x~q2:y~n:row} for both task types. In contrast, the model tends to rate designs using non-positional encodings for \texttt{q1} as negative examples. \textbf{These encoding preferences largely align with Kim et al. and common visualization design best practices.} 

While the training sets for GPT4-Turbo and Kim~\etal follow similar distributions for the value tasks, there are notable divergences, especially concerning the use of \texttt{color} and \texttt{size}, for summary tasks. Furthermore, \textbf{GPT4-Turbo displays more ``clear-cut'' preferences} than Kim~\etal's data, in that there are more encoding designs that it rates as positive or negative across the board relative to Kim~\etal This more pronounced preference may contribute to the better linear fits of LLM-provided labels than those from Kim~\etal

\subsubsection{Analyzing DracoGPT-Rank Soft Constraint Counts}
\label{sec:soft constraint count}
Many soft constraints in Draco correspond to design choices in visualizations.
For instance, \texttt{linear\_color} represents the use of a linear \texttt{color} scale, while \texttt{summary\_facet} indicates the application of faceting in a chart designed for summary tasks.\footnote{Complete descriptions for each Draco soft constraint are available at \url{https://github.com/cmudig/draco2/blob/main/draco/asp/soft.lp}.}
By analyzing how often each soft constraint is employed by positive examples versus negative examples, we can begin to map design preferences in terms of concrete visualization design decisions. 

\begin{figure*}[t]
    \centering 
    \includegraphics[height=9cm]{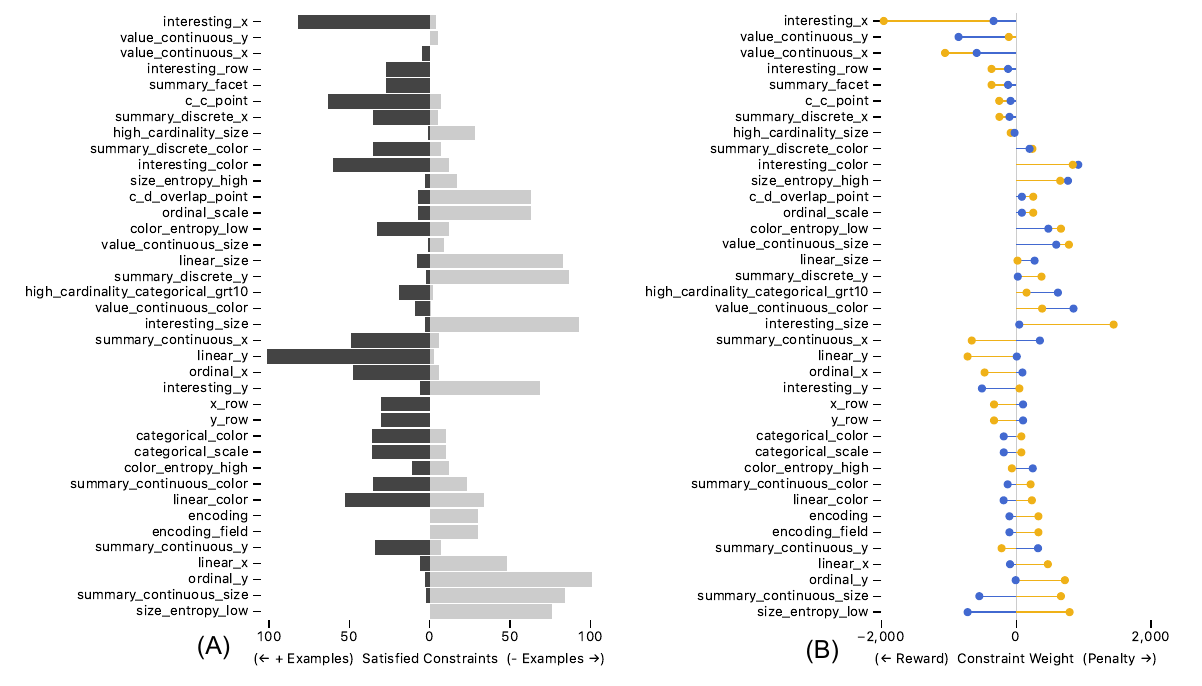}
    
    \caption{(A) The number of times each constraint is satisfied in chart pairs where GPT4-Turbo labels disagree with Kim~et~al. \add{By analyzing where \textit{constraint counts} diverge, we see, for example, that }GPT4-Turbo is more likely to label a chart negative if it uses a \texttt{size} channel (\texttt{linear\_size}, \texttt{interesting\_size}) or an \texttt{ordinal\_y} scale. (B) Constraint weights in the fitted models for Kim~\etal results (blue) and DracoGPT-Rank (gold). The weights listed last exhibit opposite signs, indicating model differences. \add{By analyzing where \textit{constraint weights} diverge, we see, for example, that} the models strongly disagree on the use of a continuous size encoding for summary tasks (\texttt{summary\_continuous\_size}).
    }
    \label{fig:preference freq}
    \vspace{-0.3cm}
\end{figure*}

% Specifically, we walk through an example of how the preferences of GPT4-Turbo and Kim et al.~diverge through the lens of soft constraint counts. F
Figure~\ref{fig:preference freq}(A) plots the frequency with which soft constraints are satisfied in positive versus negative examples from pairs for which GPT4-Turbo's responses are consistent but differ from results in Kim et al.
% Looking at the soft constraints for which GPT4-Turbo and Kim et al.~show the biggest divergence,
We see, for instance, that compared with Kim et al.'s results, GPT4-Turbo is more inclined to label charts negative if they have an ordinal \texttt{y} scale and a \texttt{size} encoding for the variable of interest, while preferring positive labels for charts with a linear \texttt{y} axis and an \texttt{x} encoding for the variable of interest.
Furthermore, Figure~\ref{fig:preference freq}(A) shows strong differences around the \texttt{y} axis scale type and the encoding channel for the variable of interest. These results hint that \textbf{GPT4-Turbo's ranking decisions may be dominated by a few design choices}, which we confirm in~\S\ref{rank weight perspective}.

\cut{To further contextualize these divergent design preferences, we compute the difference in satisfied soft constraints for each chart pair tallied in Figure(A) and identify the most common differences. 
One of the most common differences is that the positive chart satisfies \texttt{color\_entropy\_low}, \texttt{interesting\_color}, \texttt{linear\_color}, and \texttt{summary\_continuous\_color} while the negative chart satisfies \texttt{interesting\_size}, \texttt{linear\_size}, \texttt{size\_entropy\_low}, and \texttt{summary\_continuous\_size} (occurring seven times out of the 154 chart pairs). The chart pairs that exhibit this difference all resemble Figure, where GPT4-Turbo prefers color encodings over size encodings for a low-entropy quantitative variable of interest (q1). }

\subsubsection{Analyzing DracoGPT-Rank Weights}
\label{rank weight perspective}
To rank charts, Draco computes the total cost of each chart by summing the weights of all satisfied soft constraints. Thus, the weights associated with soft constraints collectively dictate Draco's design preferences. Charts with lower total costs are deemed more favorable, making it preferable to satisfy soft constraints with lower weights. Specifically, negative weights can be considered rewards as they decrease the total cost, while positive weights act as penalties.

After fitting Draco(GPT) models, we observe that most of the weights in both models are zero, indicating the models do not have preferences for these design choices. This finding echoes Zeng et al.~\cite{zeng2023too}, who also find many zero weights when training Draco instances on chart pairs drawn from graphical perception studies. Figure~\ref{fig:preference freq}(B) compares the weights of DracoGPT-Rank with Draco trained on the Kim~\etal results. Given the large number of constraints with opposite signs in the two models (constraints in the lower half of the chart), we further confirm that \textbf{GPT4-Turbo ranking preferences materially differ from Kim et al.'s results}. For instance, while GPT4-Turbo strongly penalizes a continuous \texttt{size} encoding for summary tasks (\texttt{summary\_continuous\_size}), Kim et al. heavily rewards it.

In \S\ref{sec:soft constraint count}, we noted the possibility of a few design decisions, especially the \texttt{y} axis type and the encoding choice for the variable of interest, significantly swaying GPT4-Turbo's ranking results. Our weight analysis corroborates this observation: DracoGPT-Rank tends to have more extreme weights than Kim~\etal, such as for \texttt{interesting\_x}, \texttt{interesting\_size}, and \texttt{ordinal\_y}, encouraging Draco to fixate on a few dominant design choices.
In other words, compared with experimental results, \textbf{GPT4-Turbo may focus on a few design choices while failing to attribute appropriate importance to other choices}.

\begin{figure}[t]
    \centering
    % First minipage for the first image
    \begin{minipage}{0.48\columnwidth}
        \includegraphics[width=\linewidth]{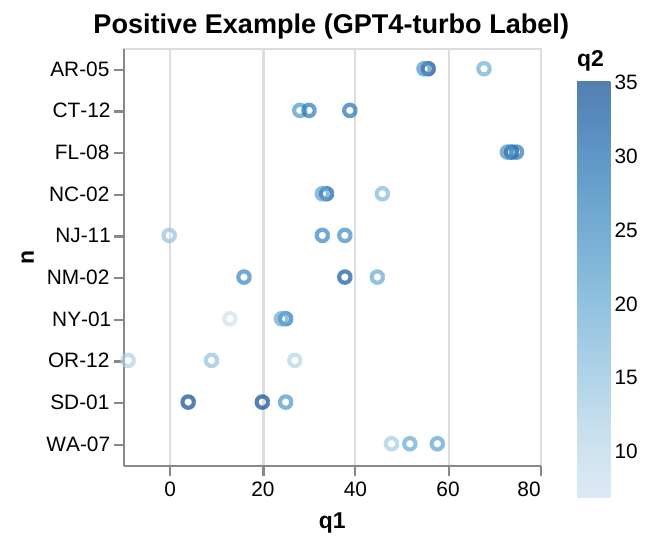} 
        \label{fig:image1}
    \end{minipage}
    \hfill 
    \begin{minipage}{0.48\columnwidth}
        \includegraphics[width=\linewidth]{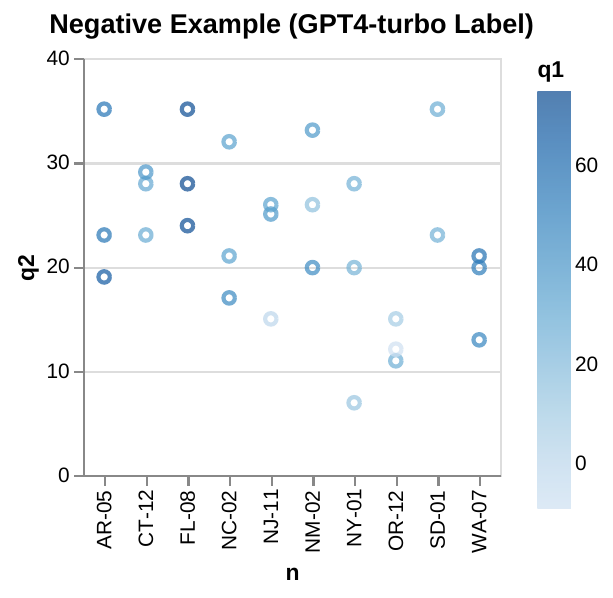} 
    \end{minipage}
    \vspace{-0.4cm}
    \caption{A chart pair demonstrating constraint weight trade-offs. Despite higher weights for a linear \texttt{x} scale (weight = 0.469) and ordinal \texttt{y} scale (weight = 0.721), GPT4-Turbo prefers the chart on the left. 
    This chart obtains a lower cost because DracoGPT-Rank has a stronger preference for encoding the variable of interest \texttt{q1} with the \texttt{x} channel (weight = -1.964) and a continuous \texttt{x} scale (weight = -1.055) for value tasks. \add{Therefore, it is important to evaluate design preferences at the chart level to complement weight-level analysis.}}
    \label{fig:pos-neg-gpt4}
    \vspace{-0.3cm}
\end{figure}

While Draco enables examination of design preferences in the form of soft constraint weights, we caution against interpreting individual weights out of context.
Even though a design element might be preferred over another in isolation, full chart designs often involve balancing additional trade-offs. 
All things being equal, DracoGPT-Rank prefers an ordinal \texttt{x} scale to a linear one and a linear \texttt{y} scale to an ordinal one.
However, Figure~\ref{fig:pos-neg-gpt4} provides an example where DracoGPT-Rank considers the chart with linear \texttt{x} and ordinal \texttt{y} scale superior to one with an ordinal \texttt{x} and linear \texttt{y} scale. 
In this case DracoGPT-Rank has an even stronger preference for encoding the variable of interest \texttt{q1} with \texttt{x} rather than \texttt{color}, and for using a continuous \texttt{x} scale over a continuous \texttt{color} scale for value tasks.
This example highlights the importance of evaluating design preferences at the chart level, not just individual weights.
In \S\ref{chart scorer} we further analyze total chart costs, comparing across models for both rank and recommend tasks.

\section{Case Study using DracoGPT-Recommend}

We now extend our case study to include DracoGPT-Recommend, \add{with a focus on examining LLM design preferences for visualization recommendation tasks and comparing them to both LLM ranking preferences and empirical findings}.
As different tool communities may have different examples and conventions, it is possible that LLMs may exhibit tool-specific design preferences.
We assess the feasibility of using DracoGPT to gauge such preferences by fitting recommendation models based on either Vega-Lite JSON or Vega-Altair Python code.

\label{recommend case study}

\subsection{Experiment Setup}
We begin with the same dataset of chart pairs from Kim~\etal For every pair in the dataset, we extract all common parts of the charts to form a partial specification. As both charts in a pair depict the same data, data facts are always preserved. Any additional commonalities in design, such as encoding channels for variables or scale types, are included in the partial specification. This process yields 1,152 partial specifications, which we represent using Draco's dictionary format (\S\ref{sec:chart-spec}). 

Next, we request optimal completions for partial specifications in two formats, Vega-Lite and Vega-Altair~\cite{vanderplas2018Altair}, which allows us to investigate whether LLM preferences vary based on the format of the output visualization specification. For all completions generated by LLMs, the pipeline discards those that are invalid due to syntactical errors or that contravene the partial specification. Each valid completion serves as the positive example in a chart pair, while charts in the original chart pair serve as negative examples. Thus, for each LLM recommendation, we can construct up to two chart pairs. The pipeline omits output pairs where the LLM-produced chart matches the original (ostensibly negative example) chart.
We use the same LLMs and train-test setup as with DracoGPT-Rank. 

\subsection{Prompts}

Our Recommend prompts follow a similar structure as our Rank prompts.
The supplemental materials include an example recommendation prompt that requests a Vega-Lite completion of a partial specification in support of a value comparison task.
We again conducted experiments to ensure instruction following before finalizing prompts.

\subsection{Fitted Draco Models}
\add{We developed a script to map Vega-Lite and Vega-Altair charts to Draco's chart representation DSL for this case study. We first validate the LLM-generated Vega-Lite specifications using \texttt{altair.Chart.from\_dict()} and the Vega-Altair specifications using \texttt{altair.Chart.to\_json()}. We include custom checks to ensure all requirements in the prompt are met, such as verifying that the charts encode all three variables.} Table~\ref{tab:valid rec} shows the proportions of valid completions of partial specifications in Vega-Lite and Vega-Altair by the three LLMs tested. Notably, GPT4-Turbo generates the highest proportion of valid completions in both formats with well over 90\% validity rates. Conversely, GPT3.5-Turbo struggles to provide valid completions for both formats, generating valid Vega-Lite completions only 19.10\% of the time and failing to produce \emph{any} valid Vega-Altair completions. Despite being explicitly prompted for Vega-Altair code, GPT3.5-Turbo persistently outputs Vega-Lite syntax. 
Common issues leading to invalid completions include syntactically incorrect code (e.g., including a ``zero'' field for the scale definition of an encoding channel for a categorical variable), violating partial specifications, and failing to encode certain variables. 

We use LLM completions as positive examples and charts from the Kim et al.~datasets as negative examples to construct chart pairs and remove pairs where the positive example is the same as the negative example. Table~\ref{tab:removal} shows the proportion of chart pairs removed for each of the training sets. In all cases, the vast majority of pairs are retained.

Table~\ref{tab:rec accuracies} shows the accuracies of the five DracoGPT-Recommend models.
% We omit the model trained on GPT3.5-Turbo Altair output because GPT3.5-Turbo fails to produce Vega-Altair completions.
Overall, average cross-validation accuracies and test accuracies are high, indicating that \textbf{DracoGPT-Recommend instances fit the design preferences expressed by their source LLMs well.}

\subsection{Design Preferences in DracoGPT-Recommend}
We now expand our analysis of LLM design preferences to include DracoGPT-Recommend. We apply the same strategies used to analyze DracoGPT-Rank results, again focusing on models fit to GPT4-Turbo responses given its superior performance.
We first inspect the encoding specification choices recommended by GPT4-Turbo. We then analyze soft constraint weights and compare DracoGPT-Rank, DracoGPT-Recommend, and Kim~\etal results. Finally, we examine total chart costs to quantify the extent to which Kim~\etal results, GPT4-Turbo's ranking preferences, and recommendation preferences align.
% In DracoGPT-Recommend names, whenever the LLM name is omitted, we are referring to GPT4-Turbo.  % We will also abbreviate Draco-Kim-Full-Set as Draco-Human.

\begin{table}[t]
  \scriptsize
  \centering
  \begin{tabu}{l*{2}{c}}
    \toprule
    & {Vega-Lite Valid} & {Vega-Altair Valid} \\
    & {Completion Rate} & {Completion Rate} \\
    \midrule
    GPT4-Turbo & 92.80\% & 95.23\% \\
    GPT4 & 88.72\% & 84.55\% \\
    GPT3.5-Turbo & 19.10\% & 0.00\% \\
    \bottomrule
  \end{tabu}
  \caption{The percentages of valid completions of partial specifications in Vega-Lite and Vega-Altair by GPT4-Turbo, GPT4, and GPT3.5-Turbo.}
  \label{tab:valid rec}
  \vspace{0.3cm}
% \end{table}
% \begin{table}[t]
  \scriptsize
  \centering
  \begin{tabu}{l*{2}{c}}
    \toprule
    & {Proportion of Pairs} & {Proportion of Pairs} \\
    & {Removed (Vega-Lite)} & {Removed (Vega-Altair)} \\
    \midrule
    GPT4-Turbo & 22.17\% & 22.97\% \\
    GPT4 & 22.70\% & 23.20\% \\
    GPT3.5-Turbo & 10.68\% & N/A \\
    \bottomrule
  \end{tabu}
  \caption{The percentages of chart pairs removed because the positive and negative examples are the same. In these cases, LLMs generate a chart that matches the original positive example from Kim et al. GPT3.5-Turbo does not produce any valid Vega-Altair completions.}
  \label{tab:removal}

  \vspace{0.3cm}
% \end{table}
% \begin{table}[t]
  \scriptsize
  \centering
  \begin{tabu}{%
      l%
        *{3}{c}%
    }
    \toprule
    & {Chart} & {Average CV} & {Test} \\
    & {Format}  & {Accuracy} & {Accuracy} \\
    \midrule
    GPT4-Turbo & Vega-Lite & 97.67\% & 96.70\% \\
    GPT4 & Vega-Lite & 99.29\% & 100.00\% \\
    GPT3.5-Turbo & Vega-Lite & 97.14\% & 96.20\% \\
    GPT4-Turbo & Altair & 98.15\% & 97.34\% \\
    GPT4 & Altair & 97.99\% & 98.67\% \\
    \bottomrule
  \end{tabu}
  \caption{Average cross-validation accuracies and test set accuracies for five DracoGPT-Recommend models. We measure the degree to which the DracoGPT instances' ranking judgments agree with the training data.}
  \label{tab:rec accuracies}
  \vspace{-0.3cm}
\end{table}

\subsubsection{LLMs' Preferred Encoding Choices}
\label{rec encodings}
\begin{figure}[t]
    \centering 
    \includegraphics[width=\columnwidth]{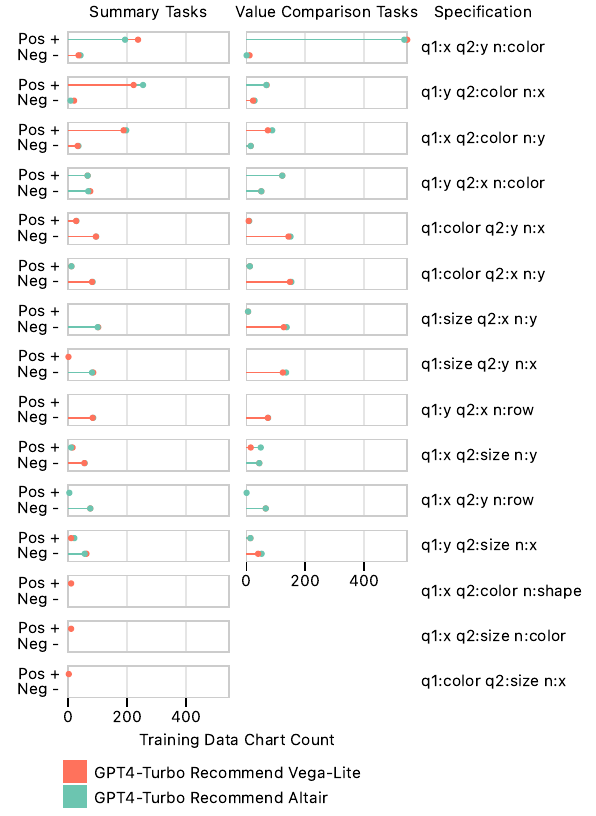}
    \caption{Distribution of positive and negative examples by encoding specification and interpretation task type for DracoGPT-Recommend chart pairs, using GPT4-Turbo to produce either Vega-Lite or Altair code.
    \add{The similarity in these training sets indicate that} encoding preferences are strongly aligned across these chart formats \add{for GPT4-Turbo.}\cut{, though both conditions produce some unique specifications.}}
    \label{fig:rec encodings}
    \vspace{-0.5cm}
\end{figure}

%Let us examine the encoding choices used by LLMs and contrast them with what negative examples use. 
Figure~\ref{fig:rec encodings} plots the distribution of encoding choices used by positive examples (GPT4-Turbo completions) and negative examples (charts from Kim et al.). Both the Vega-Lite and Vega-Altair completions produced by GPT4-Turbo largely follow the same distribution across different encoding choices, \textbf{indicating a high level of agreement in design preferences across output formats} in this case. Notably, \texttt{q1:x~q2:y~n:color} is the most common choice by GPT4-Turbo. In contrast, encoding \texttt{q1}, the variable of interest, with \texttt{color} or \texttt{size} is quite rare, which aligns with its ranking preferences established in \S\ref{rank enc}. Interestingly, GPT4-Turbo uniquely generates a few recommendations that do not exhaust positional channels before using other channels, such as \texttt{q1:x~q2:color~n:shape}, which are generally considered less effective designs~\cite{franconeri2021science}.
These examples point to potential variance in LLM-generated visualization recommendations.

\subsubsection{Analyzing DracoGPT-Recommend Weights}
\label{rec weight perspective}

\begin{figure}[t!]
  \includegraphics[width=\columnwidth]{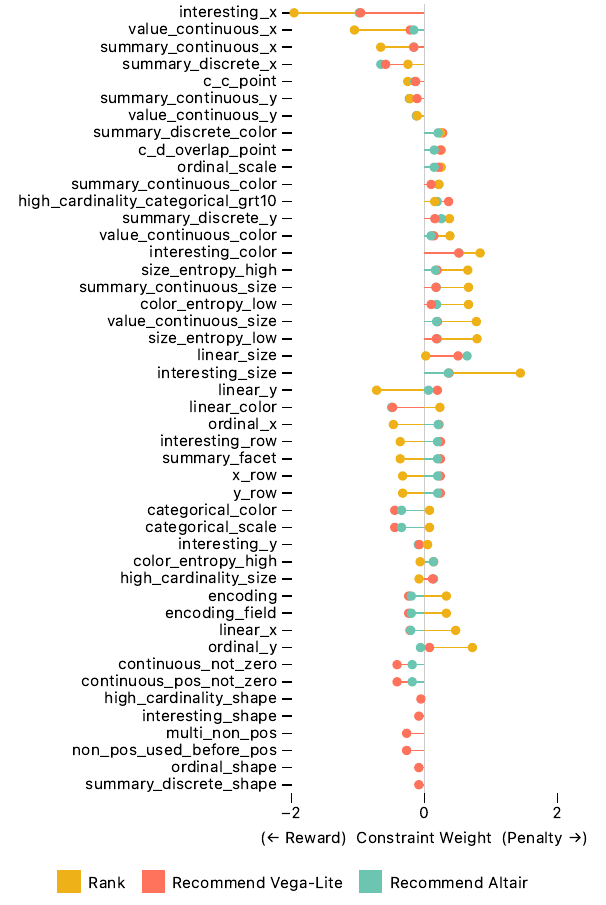}  
  \caption{Constraint weights in the fitted GPT4-Turbo DracoGPT models for Rank, Recommend Vega-Lite, and Recommend Altair. The weights \add{indicate} highly aligned recommendation preferences for Vega-Lite and Altair, while both differ from ranking preferences.}
  \label{fig:rec weights}
  \vspace{-0.3cm}
\end{figure}

After training DracoGPT-Recommend, we can again analyze soft constraint counts and weights to examine fine-grained design preferences. Here, we illustrate what the weights reveal about design preferences encoded by DracoGPT-Rank and DracoGPT-Recommend. Figure~\ref{fig:rec weights} depicts the constraint weights for Draco instances fit to Vega-Lite recommendations, Altair recommendations, and our earlier rank labels (from \S\ref{rank weight perspective}). 
Figure~\ref{fig:rec weights} shows that some constraints have non-zero weights for Recommend Vega-Lite, but weights of zero for the other two models. The Vega-Lite visualizations generated by GPT4-Turbo cover a wider design space and satisfy a wider set of constraints than the charts from Kim~\etal or GPT4-Turbo Altair recommendations.
In such cases, the weights for these constraints may be treated skeptically, due to the lack of design space coverage of negative examples. Among the 1,069 valid recommendations by GPT4-Turbo, only ten utilize the \texttt{shape} channel yet all \texttt{shape}-related constraints are linked to rewards. As none of Kim~\etal's charts use a \texttt{shape} encoding, there are no potentially offsetting negative examples.

Judging by the quantity of constraints with different signs, it becomes evident that \textbf{GPT4-Turbo's design preferences for recommendation, across both output formats, align more closely with each other than with GPT4-Turbo's design preferences for ranking}.
The only divergence between the Recommend Vega-Lite and Recommend Altair models is in whether an ordinal \texttt{y} scale is preferable.
Meanwhile, both of these models disagree with DracoGPT-Rank on 15 design choices. For example, we see that when ranking, GPT4-Turbo favors faceting, but penalizes it in a recommendation context.
This divergence stems from the fact that GPT4-Turbo generally labels faceted charts positive, but rarely generates them when prompted for recommendations.

\subsubsection{Assessing Alignment via Chart Cost Correlations}
\label{chart scorer}
As discussed in \S\ref{rank weight perspective}, design choices do not exist in a vacuum. A good design element in one chart may be undesirable in another. Therefore, it is helpful to complement constraint-level analysis with chart-level analysis to account for interactions among design choices. 
Given an input chart, Draco computes a total cost by summing the weights of all satisfied constraints: the lower the cost, the more preferable the chart.
As two Draco instances with similar design preferences should produce similar patterns of costs over a collection of charts, the product-moment correlation among costs can serve as a measure of model alignment.

Accordingly, we extract all distinct chart stimuli from Kim et al.~and compare their costs as evaluated by various Draco instances. We consider two charts distinct if their feature vectors (soft constraint satisfactions) differ.
The Kim~\etal stimuli consist of 524 different charts, which map to 48 distinct feature vectors.
Many charts use the same visual specification for different input data; however, the data features (such as coarsely-binned entropy values) may map to the same Draco representations.
Thus we avoid double counting charts with identical designs with data sampled from similar distributions, which would unduly affect measured correlation strengths.

\begin{figure}[t]
    \centering 
    \includegraphics[width=\columnwidth]{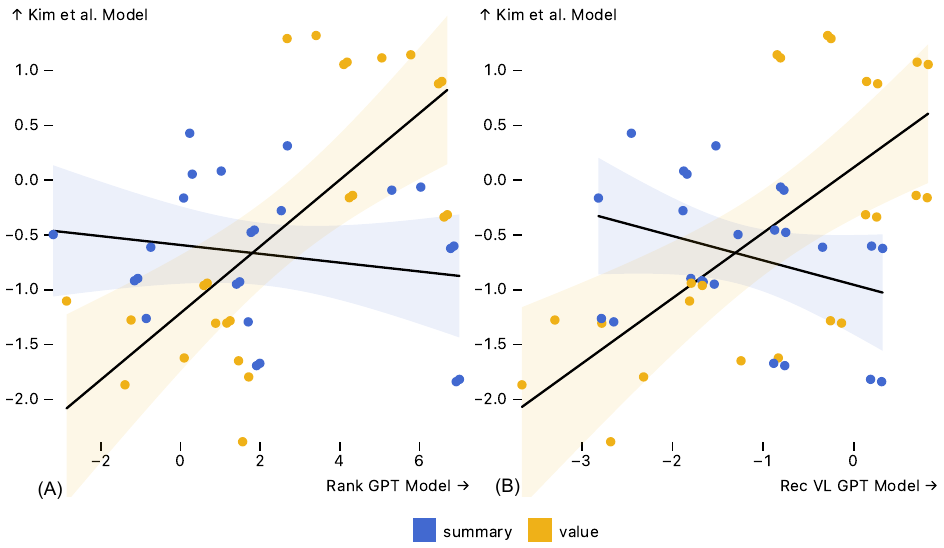}
    \caption{\add{Chart costs conditioned on task type, comparing Kim~\etal's experimental results
    to DracoGPT-Rank (A) and DracoGPT-Recommend Vega-Lite (B).
    For value tasks, results from both GPT4-Turbo models moderately correlate with a model fit to human performance data, but do not significantly correlate for summary tasks.}}
    \label{fig:kim_vs_combined_by_task}
\end{figure}

Figure~\ref{fig:kim_vs_combined_by_task}(A) presents the costs according to GPT4-Turbo Rank and Kim~\etal's experimental results\add{, conditioned on perceptual (summary vs. value) task}, while Figure~\ref{fig:kim_vs_combined_by_task}(B) shows the costs according to GPT4-Turbo Recommend Vega-Lite and the Kim~\etal results\add{, again conditioned on task}. Pooling across perceptual tasks, the costs from a Draco model fit to Kim~\etal results have a weak positive correlation with both DracoGPT-Rank ($r(46) = 0.36$, $p = 0.011$) and DracoGPT-Recommend Vega-Lite ($r(46) = 0.40$, $p = 0.005$), which suggests that \textbf{empirical results from Kim et al. do not align well with either GPT4-Turbo's ranking or recommendation preferences}.

\add{Conditioning on task type, we see that Kim~\etal Draco model costs have a moderately strong positive correlation with both DracoGPT-Rank ($r(22) = 0.69$, $p < 0.001$) and DracoGPT-Recommend Vega-Lite ($r(22) = 0.67$, $p < 0.001$) for the value task. For the summary task, however, they exhibit negative correlations that are not statistically significant with either DracoGPT-Rank ($r(22) = -0.18$, $p = 0.408$) or DracoGPT-Recommend Vega-Lite ($r(22) = -0.32$, $p = 0.127$), indicating a specific divergence between GPT4-Turbo preferences and Kim~\etal results for summary judgments. Though it is difficult to pinpoint the exact reason for the poor alignment of the model's preferences and empirical results, we conjecture that the LLM responses may reflect common visualization patterns in LLM training data. As much visualization research and discourse focus on value tasks~\cite{kim2018assessing}, the next-token pretraining objective may incentivize LLMs to prioritize encoding information relevant to value tasks.}

\begin{figure}[t]
    \centering 
    \includegraphics[width=\columnwidth]{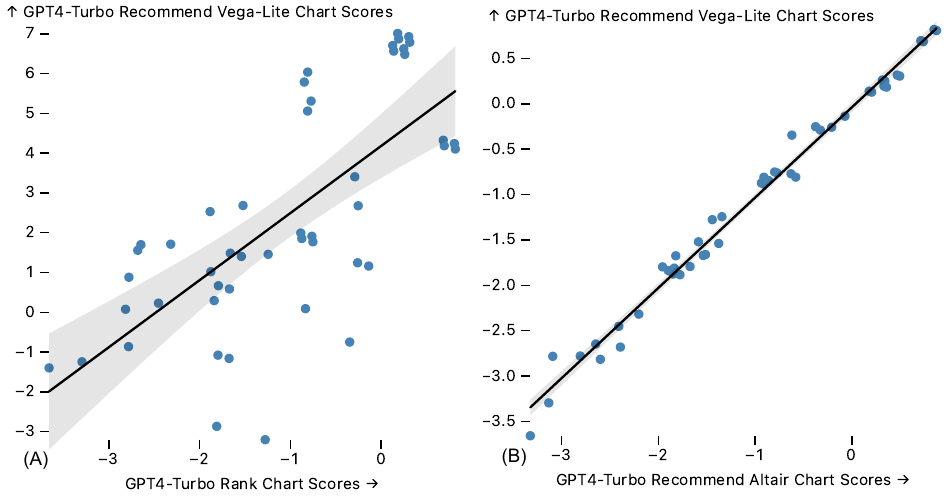}
    \caption{(A) Chart costs according to DracoGPT Rank and Recommend Vega-Lite models; (B) Chart costs according to DracoGPT Recommend Altair and Recommend Vega-Lite models. GPT4-Turbo's Rank and Recommend Vega-Lite preferences exhibit moderate correlation, while chart scores from Vega-Lite and Altair recommendations strongly correlate.}
    \label{fig:rec_vl_vs_combined}
    \vspace{-0.3cm}
\end{figure}

Figure~\ref{fig:rec_vl_vs_combined}(A) presents the costs according to DracoGPT-Rank and GPT4-Turbo Recommend Vega-Lite. These costs have a moderately strong positive correlation ($r(46) = 0.70$, $p < 0.001$). This suggests that \textbf{GPT4-Turbo’s ranking and recommendation preferences align moderately well with each other}. In other words, GPT4-Turbo is more consistent with itself across these two tasks than it is with Kim~\etal's empirical results. These findings reinforce our contention that ranking and recommending are related yet distinct visualization tasks that warrant separate investigation.

Figure~\ref{fig:rec_vl_vs_combined}(B) presents the costs according to GPT4-Turbo Recommend Vega-Lite and -Altair. The chart costs exhibit a near-perfect positive correlation ($r(46) = 0.99$, $p < 0.001$). This indicates that \textbf{GPT4-Turbo exhibits highly consistent design preferences when making either Vega-Lite or Vega-Altair recommendations}, likely due to the high-level of similarity between the two formats.

\section{Discussion: Limitations \& Future Work}
In this paper, we present DracoGPT, a method for extracting, modeling, \add{and assessing} visualization design preferences from LLMs and evaluate them using the Draco knowledge base. Motivated by different use cases of LLMs for data visualization, we propose to extract design preferences of LLMs for two tasks: ranking visualization pairs and recommending visualizations.
Using data from Kim et al.~\cite{kim2018assessing} and three LLMs as a case study, we show through high test set accuracies that both DracoGPT-Rank and -Recommend produce models that accurately match the design preferences expressed by LLMs.
To analyze the design preferences modeled by DracoGPT, we compare the visual encoding choices used by positive and negative training data examples, tally the soft constraints that positive and negative examples tend to satisfy, inspect learned soft constraint weights, and examine correlations among total chart scores.
Each unit of analysis provides complementary perspectives for assessing the alignment between LLMs and empirical graphical perception data. We show, for example, that while GPT4-Turbo exhibits some ranking preferences akin to those observed in Kim et al., many of its preferences differ.
When using Draco to score charts according to extracted preferences, we find that GPT4-Turbo's ranking and recommendation preferences are moderately aligned, \add{while both are moderately aligned with findings from Kim et al. for the value task and misaligned for the summary task.}

One limitation of our case study is the limited visualization design space covered by the Kim \etal~dataset. Though covering a range of encoding specifications, all the charts are scatter plots depicting two quantitative fields and a nominal field. Utilizing the DracoGPT pipeline, future work can extract a wider set of visualization design preferences by expanding the design space covered by chart pairs. Zeng~et~al.~\cite{zeng2023too} compiled a dataset of chart pairs from 30 studies and mapped them to Draco representations, providing a useful resource for future endeavors. \add{To enable a comprehensive understanding of LLM visualization design preferences, we further call upon future graphical perception studies to publish their stimuli.}

In addition, future work should expand Draco’s knowledge base. By adding more soft constraints representing a wider range of design decisions (e.g., if a visualization uses grid lines), we can better model more nuanced design preferences from LLMs. In order to identify shortcomings of Draco, we can generate a wide array of chart pairs using LLMs in a semi-controlled manner and check if Draco is able to discriminate them. Draco's failure to do so may reveal design decisions that its knowledge base cannot encode.

As our case study found that DracoGPT can accurately model LLM preferences, in the future Draco may have the potential to serve as a reliable and cost-effective stand-in for LLMs. Eliciting visualization recommendations from LLMs is compute- and cost-intensive, whereas Draco is a lightweight model that is cheap to deploy.
If Draco's constraint set is suitably expanded and trained on chart pairs covering a large design space, it could serve as a valuable and efficient means of reifying and reusing preferences implicit to an LLM.

DracoGPT also provides a means for studying the influence of input and output formats on the visualization design preferences expressed by an LLM.
In this work, we present input visualizations to LLMs using Draco's dictionary format.
Future work can explore alternative specifications, including images.
Furthermore, we hope to request recommendations in formats other than Vega-Lite and Vega-Altair, such as the popular ggplot2~\cite{wickham2016ggplot2} and Matplotlib~\cite{hunter2007matplotlib} libraries.
We found little difference in GPT4-Turbo design preferences when recommending Vega-Lite JSON or Vega-Altair Python code. However, this result is not particularly surprising given the close relationship between these tools (Altair is a Python API for Vega-Lite) and overlapping online examples.
Other visualization tools, through their differing communities, use cases, and examples, may lead to differing LLM design preferences that can be analyzed via DracoGPT pipelines.
While the engineering involved to transform other specifications into a form amenable to Draco is non-trivial, reverse-engineering tools such as DIVI~\cite{snyder2023divi} raise the possibility of leveraging SVG as a common medium for translating charts into a Draco-compatible format.

We also require further research on how users interact with LLMs for visualization-related tasks. Here we proposed two tasks (ranking and recommendation) that users can perform with LLMs, and there are undoubtedly more, such as critiquing visualization designs. A better understanding of such tasks could lead to future work enabling DracoGPT to extract visualization design preferences in more contexts.

Finally, we presented both low-level and high-level measures of alignment of design preferences between different sources (e.g., empirical research, LLMs, and personal preferences).
Future work could leverage DracoGPT to expose how LLM design preferences differ in response to techniques such as fine-tuning or retrieval-augmented generation (RAG). To this end, DracoGPT can serve as a tool to evaluate the results of various attempts at LLM alignment in support of more effective visualization tasks.

\section*{Supplemental Materials}

In Supplemental Material 1, we provide example LLM prompts for both DracoGPT-Rank and DracoGPT-Recommend. In Supplemental Material 2, we provide queries and responses for all DracoGPT runs. The folder named ``rank'' contains data for DracoGPT-Rank. The folder named ``recommend'' contains data for DracoGPT-Recommend.

\acknowledgments{
We thank the reviewers for their helpful feedback. This work was partially supported by the NSF (award \# IIS-2141506) and an OpenAI Research Grant.
}

\bibliographystyle{abbrv-doi-hyperref}
\bibliography{template}

\end{document}